# Coaxial Microwave Plasmas in Argon: Radial Contraction, Self-Organization and more Exotic Phenomena

E.A.D. Carbone[1,2], S. Hübner, E. M. van Veldhuizen, A. Schrader, S. Nijdam, G.M.W. Kroesen

*Abstract* – A coaxial microwave plasma setup was designed for investigation by optical and laser diagnostics. The plasma is sustained by two microwave power sources located at both ends of a coaxial line. This allows generating an axially homogeneous glow discharge at low pressure. For increasing pressures, this glow-like mode is found to be unstable and various self-organized patterns are observed including finger-like structures which can be sustained after the visible end of the plasma column.

A low pressure microwave plasma in coaxial configuration is generated using two microwave power (2.45 GHz) generators located at both ends of a 1 m coaxial line where the plasma behaves as the outer conductor (see figure 1.f for a schematic view). This type of plasma is widely used for deposition on large areas of films such as graphene [1]. In this study, pure argon gas is fed into the system and the pressure is varied while changing the flow and/or pumping speed. The power can be increased from 100 W up to 2 kW from both magnetrons.

In figure 1.a, an optical picture of the plasma generated around the quartz tube is shown for a pressure of 0.14 mbar. One can see that the plasma is both axially and radially diffuse. The plasma light intensity peaks close to the tube and monotonously decreases towards the chamber wall while only a weak gradient can be observed along the z-axis.

For increasing pressures, a contraction of the radial light emission intensity profile is observed. This is illustrated in figure 1.b for a pressure of 3.8 mbar. One can see that the plasma intensity peaks close to the wall and it is surrounded by a more diffuse plasma which ends a couple of centimeters away from the quartz. This can be explained by a competition between the local ionization and losses and transport of electrons but also energy transport in the radial direction [2]. There is a transition from an ionizing to a recombining plasma region in front of the wall.

At a pressure around 5 mbar, a transition to a different plasma mode occurs. The azimuthal symmetry of the plasma is broken and one can see 'filamentary' structures (stripes) parallel to the z-axis and for higher pressures structures interconnections between the stripes are observed (figure 1.c). These structures are very reminiscent of the self-organized structures seen by various authors in DBD discharges [3]. The formation of these periodic structures is usually explained by the non-linear coupling between production and transport processes which leads to Turing patterns formation [4]. Similarly, for higher power, (see fig. 1.d), the axial homogeneity of the filaments is broken and a succession of small plasma balls can be seen. These axial striations can be correlated to the same type of instabilities as discussed for the bifurcation towards the horizontal stripes pattern. Similar structures were observed in the past for other types of microwave discharges [5].

In the high pressure regime, a new kind of self-organization is observed (figure 1.e). Finger-like structures standing on the quartz tube are formed towards the end of the plasma column and remain stable beyond the end of the plasma column or between the two columns when both magnetrons are used. These structures are found to generally move away from the plasma region where they were created and to go upwards, the latter likely due to convection. The mechanism of formation of these structures is still a mystery but one could note that the microwave self-shielding properties of the plasma are lost under these conditions. These structures might then be partially sustained by radiating EM energy.

To conclude, one can see that the use of 2D self-consistent models is not sufficient any longer when the pressure is increased above a few mbar. A full 3D description is needed, and considering that the frequency spectrum of the magnetrons contains various lower frequencies, time dependence might have to be further elaborated as well.

Manuscript received XXX 2013; revised XXX 2014.
The authors are with the Technical University of Eindhoven, Department of Applied Physics, Den Dolech 2, 5600MB Eindhoven, the Netherlands
[1] Currently at CNRS-LTM, rue des Martyrs 17, 38000 Grenoble, France, [2] Corresponding author: emile@epgmod.phys.tue.nl
Work was supported by the Dutch Technology Foundation (STW) under the project numbers 10744 and 10497.
 Publisher Identifier S XXXX-XXXXXXX-X

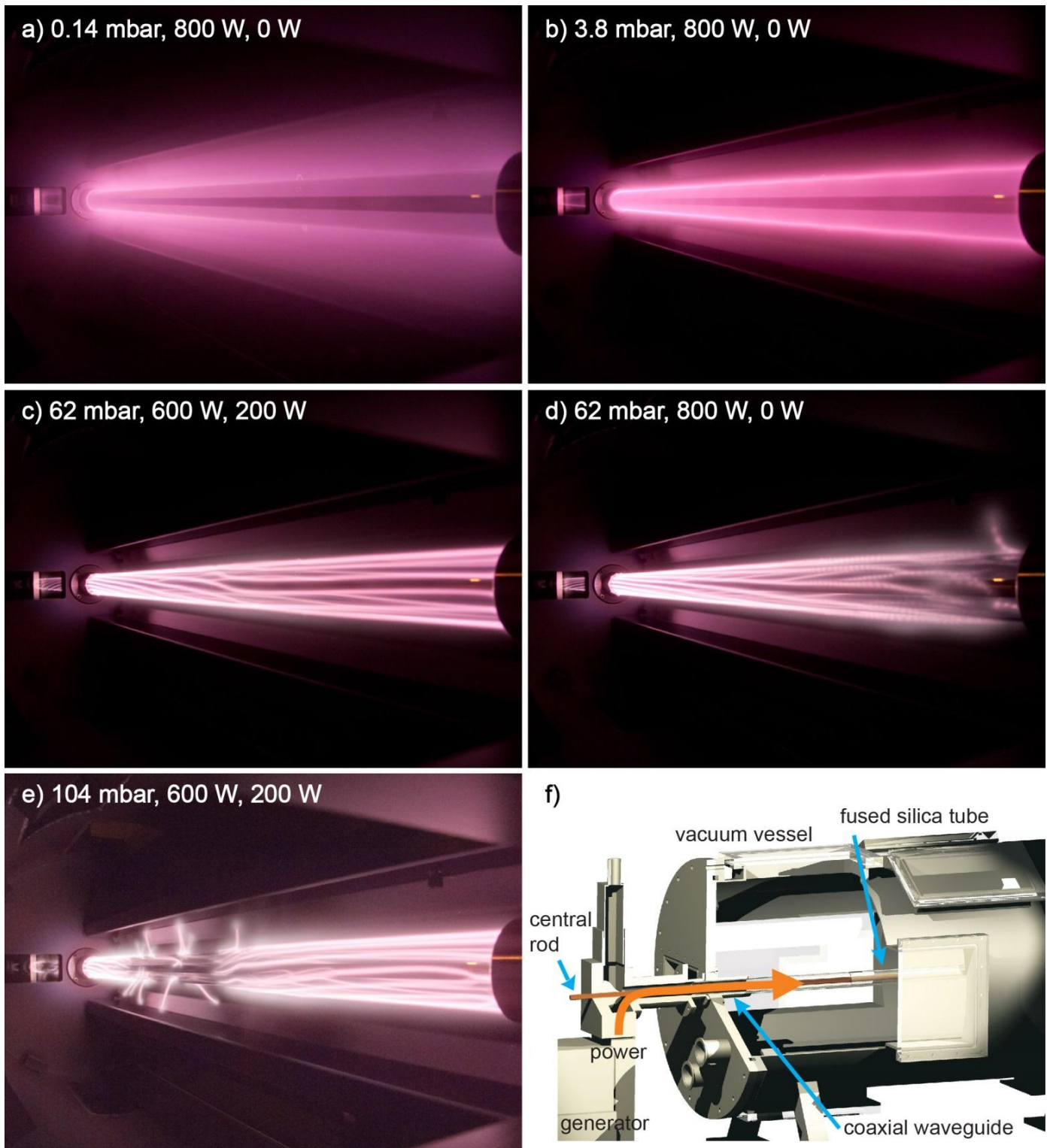

Fig. 1. a)-e) digital camera images of the coaxial plasma set-up operated in an argon atmosphere at varying pressures and applied microwave powers as indicated in the figure. The indicated powers are for the left (most distant) and right (closest) magnetron respectively. Note that camera aperture and exposure values vary between images. f) Schematic drawing of the vacuum chamber, coaxial structure, one magnetron and the direction of the power flux.